\documentclass[twocolumn,aip,jcp,superscriptaddress,citeautoscript,longbibliography,floatfix,reprint]{revtex4-2}
\usepackage{graphicx}
\usepackage{amsmath}
\usepackage{amssymb}
\usepackage{comment}
\usepackage{color}

\newcommand{\tn}[1]{\textnormal{#1}}    
\newcommand*{\Na}{N_{\tn{A}}}           
\newcommand*{\na}{n_{\tn{A}}} 
\newcommand*{\nb}{n_{\tn{B}}} 

\newcommand*{\lb}{l_{\tn{B}}}
\newcommand*{\pH}{\tn{pH}}
\newcommand*{\pOH}{\tn{pOH}}
\newcommand*{\pKa}{\tn{p}K_{\tn{a}}}
\newcommand*{\pKb}{\tn{p}K_{\tn{b}}}

\newcommand*{\nam}{n_{\tn{A}^-}}        
\newcommand*{\nbp}{n_{\tn{B}^+}}        
\newcommand*{\nxp}{n_{\tn{X}^+}}        
\newcommand*{\nxm}{n_{\tn{X}^-}}        
\newcommand*{\muxp}{\mu_{\tn{X}^+}}     
\newcommand*{\muxm}{\mu_{\tn{X}^-}}     
\newcommand*{\pI}{\tn{pI}}              
\newcommand*{\pIxp}{\tn{pI}_{\tn{X}^+}} 
\newcommand*{\pIxm}{\tn{pI}_{\tn{X}^-}} 

\renewcommand{\emptyset}{\varnothing}

\usepackage{url}

\newcommand{\figurewidth}{.9\textwidth}
\newcommand{\narrowfigurewidth}{.45\textwidth}
\renewcommand{\epsilon}{\varepsilon}

\usepackage{bm}

\begin{document}

\title{Accelerated simulation method for charge regulation effects}

\author{Tine Curk}
\email{curk@northwestern.edu}
\affiliation{Department of Materials Science and Engineering, Northwestern
University, Evanston, Illinois 60208, USA}

\author{Jiaxing Yuan}
\affiliation{School of Physics and Astronomy \& Institute of Natural
  Sciences, Shanghai Jiao Tong University, Shanghai 200240, China}
 \affiliation{Research Center for Advanced Science and Technology, University
   of Tokyo, 4-6-1 Komaba, Meguro-ku, Tokyo 153-8904, Japan}

\author{Erik Luijten}
\email{luijten@northwestern.edu}
\affiliation{Department of Materials Science and Engineering, Northwestern
University, Evanston, Illinois 60208, USA}
\affiliation{Departments of Engineering Sciences \& Applied Mathematics,
  Chemistry, and Physics \& Astronomy, Northwestern University, Evanston,
  Illinois 60208, USA}

\begin{abstract}
  The net charge of solvated entities, ranging from polyelectrolytes and
  biomolecules to charged nanoparticles and membranes, depends on the local
  dissociation equilibrium of individual ionizable groups. Incorporation of
  this phenomenon, \emph{charge regulation}, in theoretical and computational
  models requires dynamic, configuration-dependent recalculation of surface
  charges and is therefore typically approximated by assuming constant net
  charge on particles. Various computational methods exist that address
  this. We present an alternative, particularly efficient charge regulation
  Monte Carlo method (CR-MC), which explicitly models the redistribution of
  individual charges and accurately samples the correct grand-canonical charge
  distribution. In addition, we provide an open-source implementation in the
  LAMMPS molecular dynamics (MD) simulation package, resulting in a hybrid
  MD/CR-MC simulation method. This implementation is designed to handle a wide
  range of implicit-solvent systems that model discreet ionizable groups or
  surface sites. The computational cost of the method scales linearly with the
  number of ionizable groups, thereby allowing accurate simulations of systems
  containing thousands of individual ionizable sites. By matter of
  illustration, we use the CR-MC method to quantify the effects of charge
  regulation on the nature of the polyelectrolyte coil--globule transition and
  on the effective interaction between oppositely charged nanoparticles.
\end{abstract}

\maketitle

\section{Introduction}
\label{sec:intro}

Acid--base ionization reactions in aqueous solutions are among the most common
chemical processes.  Many soft and biological materials, including colloidal
nanoparticles, polyelectrolytes, proteins, and membranes, acquire charge due
to ionization of acidic or basic surface groups~\cite{atkins2006}. The degree
of ionization depends on the pH and salt concentration of the solution, but
may also be strongly influenced by the presence of other charged entities in
the vicinity. This \emph{charge regulation} (CR) effect~\cite{ninham71} can
strongly enhance protein--protein~\cite{lund2005a,lund2013,roosen-runge2014}
and protein--membrane~\cite{lund2005b} interactions, reduce the electrostatic
repulsion between like-charged nanoparticles~\cite{takae2018,dosSantos2019},
and modulate the self-organized morphology of polyelectrolyte
brushes~\cite{tagliazucchi2010,barr2012}.  Moreover, CR effects can be
significantly stronger than surface polarization effects~\cite{kirkwood1952},
and its many-body nature can even qualitatively change the self-assembled
structures of charged nanoparticles~\cite{curk2021}.  Charge regulation is
also directly relevant to numerous practical applications. For example, the
response of weak polyelectrolytes to external stimuli enables the design of
ionic current rectifiers~\cite{yameen09a,guo10,Tagliazucchi13} and controlled
drug release~\cite{huang2019}.

Despite the important role of CR, theoretical and computational studies of
solvated systems still widely employ the constant charge (CC) approximation
due to the relative simplicity of its implementation. For example, constant
charges result in constant interaction potentials that are straightforward to
use in molecular dynamics (MD) simulations. Conversely, CR requires the
dynamic computation of ionization states that depend on the instantaneous
microstructure of a system, leading to structure-dependent interaction
potentials that greatly increase computational complexity and cost.  As a
result, the CC approximation is routinely used in regimes where it is not
appropriate, e.g., in partially ionized systems---a choice that is
particularly striking given the key role that electrostatic interactions play
in nanoparticle aggregation and self-assembly
processes~\cite{walker11,boles2016} and the function of
biomolecules~\cite{zhou2018}.

Theoretical efforts to accurately describe CR have been ongoing since the
1950s~\cite{kirkwood1952}, providing valuable insight into electrostatics of
membranes~\cite{ninham71,majee2019}, colloidal
interactions~\cite{trefalt2016,markovich2016,bakhshandeh2020} and
polyelectrolyte conformations~\cite{netz2002,prusty2020}, but have remained
confined to the Poisson--Boltzmann description of the electrolyte and
relatively simple geometries~\cite{podgornik2018}.  Conversely, particle-based
simulations can offer much more accurate representations and greater
versatility.  From a microscopic point of view, acid--base reactions involve
the formation and breaking of chemical bonds, which requires the use of
ab-initio MD where the interatomic forces are computed on the
fly~\cite{lu2008,maurer2010}.  However, such calculations are computationally
very costly and therefore limited to extremely short time and length
scales. By comparison, generic coarse-grained models are simpler to use and
orders of magnitude faster.

In coarse-grained simulations, two common techniques for modeling acid--base
equilibria are the constant-pH Monte Carlo (MC)
method~\cite{reed1992,radak2017} and the reaction ensemble MC (RxMC)
method~\cite{smith1994,johnson1994,turner2008}.  Both methods have been used
to model ionizable charged surfaces~\cite{barr2011}, weak polyelectrolytes in
bulk
solution~\cite{reed1992,ulrich2005a,carnal2010,narayanan2014,landsgesell2017,murmiliuk2018,whitmer2019}
and near
interfaces~\cite{ulrich2005b,ulrich2006,carnal2011,barr2012,stornes2017,stornes2018},
hydrogels~\cite{hofzumahaus2018}, and
proteins~\cite{lund2005a,lund2005b,lund2013}.  The constant-pH method treats
pH as an input parameter without explicitly considering dissociated protons.
Therefore, the method is only applicable if the monovalent salt concentration
is much higher than the concentration of dissociated ions ($\tn{H}^+$,
$\tn{OH}^-$), such that the presence of these ions can be neglected.  In
contrast, the RxMC method explicitly models dissociated ions and is thus
applicable at low salt concentrations as well.  However, the RxMC method
requires that dissociated ions are exchanged with a reservoir only in
``corresponding groups,'' where a group refers to, e.g., the ions making up a
specific salt. Consequently, the method can only exchange dissociated ions if
those also exist as a component of an additional salt. Even then, it does not
lead to the correct grand-canonical distribution of individual ions. The
approximation leads to particularly severe finite-size effects at low ion
concentrations, e.g., at $\pH \approx 7$. As a rule of thumb, the pH range
where RxMC ($\tn{pH}\le3$ or $\tn{pH}\ge11$) is acceptable is complementary to
the application range of the constant-pH method
($3\le \tn{pH}\le11$)~\cite{landsgesell2017,kosovan2019}.

We propose an improved CR-MC method that is accurate and efficient over the
full range of pH values and salt concentrations.  Our method builds on the
RxMC approach~\cite{smith1994,johnson1994,turner2008}, but consistently
implements both ionization and the exchange of individual ions, and correctly
samples the grand-canonical distribution.  Moreover, our CR framework is not
limited to acid--base reactions, but can be directly applied to any ionization
process, broadly defined, including surface adsorption of charged entities
such as salt ions or macro-ions.

Except for a coarse-grained implementation of the RxMC method in
ESPResSo~\cite{limbach2006} and an atomistic constant-pH ensemble
implementation in NAMD~\cite{radak2017}, we are not aware of any open-source
MD packages that are capable of simulating CR phenomena. To benefit the
computational community, we present an open-source, parallel implementation of
the CR-MC method in the LAMMPS MD package~\cite{plimpton95}, providing a
powerful and highly efficient MD/MC hybrid tool for modeling CR effects in
solvated systems.

Upon developing this method we became aware of the grand-reaction
approach~\cite{landsgesell2020}, which provides a general framework for
simulating chemical reactions in the grand-canonical ensemble.  When applied
to coarse-grained electrolyte models, the grand-reaction approach is
thermodynamically equivalent to our CR-MC method and leads to the same
equilibrium distribution of charged states. The main difference, however, is
that we employ a more efficient MC sampling scheme and that our implementation
is optimized specifically for charge regulation.  Taken together, these
factors result in a sampling rate that is nearly an order of magnitude faster
than the approach of Ref.~\onlinecite{landsgesell2020}, allowing us to
simulate previously intractable systems with thousands of ionizable
sites~\cite{curk2021}. Our approach can thus be seen as an optimization of the
grand-reaction ensemble method~\cite{landsgesell2020}.

This article is organized as follows. In Sec.~\ref{sec:model-algorithm}, we
present an overview of the CR model, along with a detailed derivation of the
CR-MC method, followed by a performance analysis.  In Sec.~\ref{sec:results},
we apply the method to investigate how CR affects two prototypical systems,
namely the coil--globule transition of a hydrophobic weak polyelectrolyte and
the effective interaction between oppositely charged nanoparticles with
variable surface charge density.  Lastly, the Appendices provide mathematical
derivations, implementation details, and numerical tests.

\section{Model and algorithm}
\label{sec:model-algorithm}

\subsection{Charge regulation model}

\begin{figure*}
\centering 
\includegraphics[width=\figurewidth]{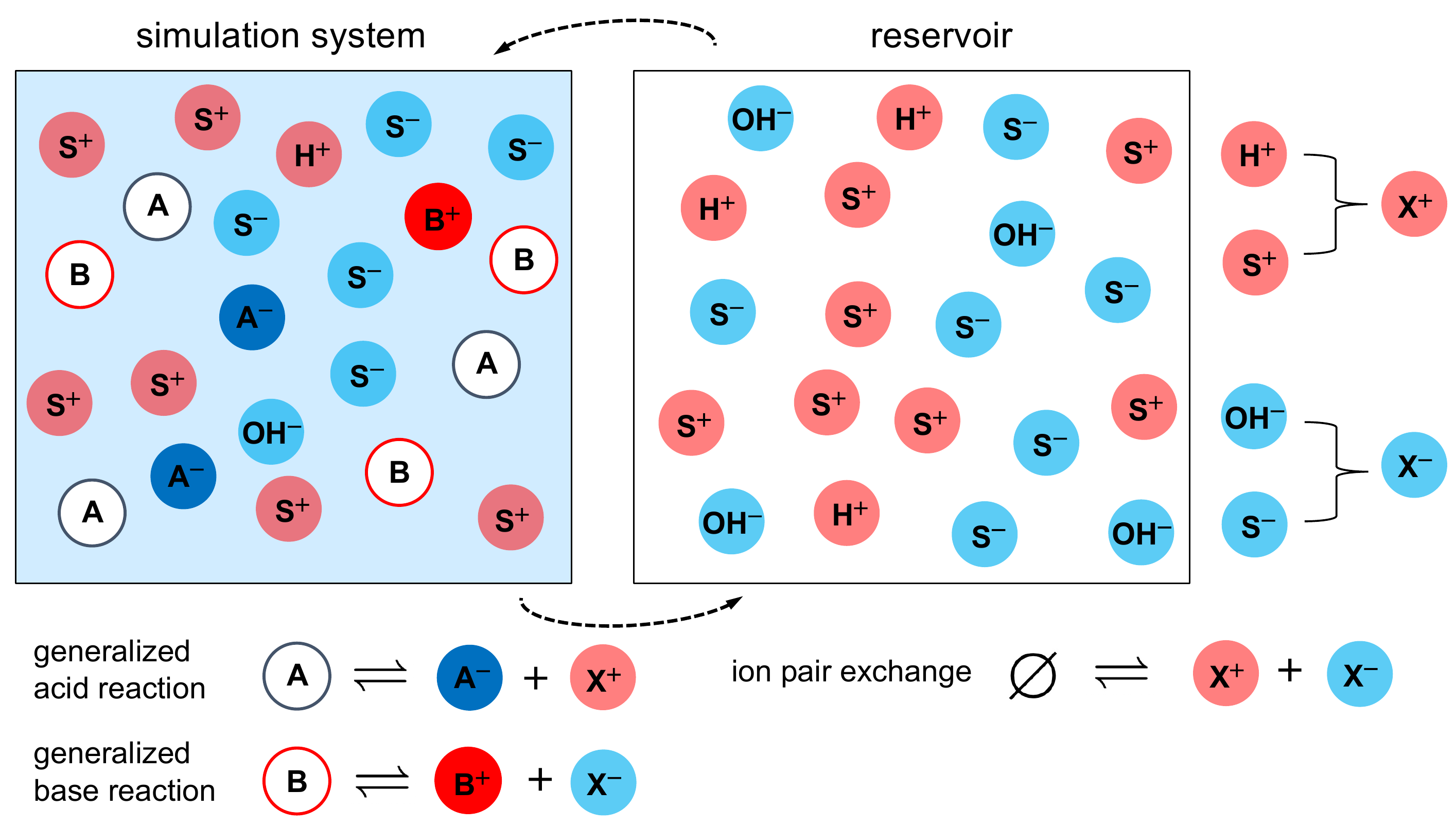}
\caption{Schematic of an acid--base reaction system in equilibrium with a
  reservoir at a fixed pH and salinity.  The acid~(A) and base~(B) groups can
  undergo ionization reactions, becoming charged (A$^-$, B$^+$) via the
  release of ions ($\tn{H}^+$, $\tn{OH}^-$). Salt cations ($\tn{S}^+$) and
  anions ($\tn{S}^-$), as well as released ions, are exchanged between the
  system and the reservoir.  In principle, all possible reactions that
  preserve the charge neutrality of the system must be considered.  Grouping
  all free monovalent ions into a single particle type (X) markedly reduces
  the number of required reactions and increases MC sampling performance.}
\label{fig:model}
\end{figure*}

We consider charged particles immersed in an implicit solvent at constant
temperature~$T$.  The particles can represent acidic groups~(A), basic
groups~(B), dissociated ions ($\tn{H}^+$, $\tn{OH}^-$), free cations
($\tn{S}^+$), and free anions ($\tn{S}^-$) (Fig.~\ref{fig:model}). The
following reactions may occur,
\begin{eqnarray}
  \tn{A} &\rightleftharpoons& \tn{A}^- +  \tn{H}^+\;,
\label{eq:acid} \\
  \tn{B} &\rightleftharpoons& \tn{B}^++\tn{OH}^-\;,
\label{eq:base}
\end{eqnarray}
where $\tn{A}^-$ and $\tn{B}^+$ denote the ionized states of the respective
groups. We also take into account the self-dissociation of water,
\begin{equation}
  \emptyset \rightleftharpoons \tn{H}^++\tn{OH}^-\;,
\label{eq:H2O}
\end{equation}
where $\emptyset$ denotes an empty set since the solvent ($\tn{H}_2\tn{O}$) is
not explicitly considered.  The system is in equilibrium with a reservoir at a
given salinity, which can be formally expressed as
\begin{equation}
  \emptyset \rightleftharpoons \tn{S}^++\tn{S}^-\;.
\label{eq:salt}
\end{equation}
A natural choice to simulate the reactions Eqs.\
(\ref{eq:acid})--(\ref{eq:salt}) is to employ the RxMC
method~\cite{smith1994,johnson1994}, which provides a framework for modeling
arbitrary chemical reactions. However, this method is limited to physical
reactions and consequently, as noted in Sec.~\ref{sec:intro}, implementing
Eqs.\ (\ref{eq:acid})--(\ref{eq:salt}) alone does not generally lead to the
correct grand-canonical distribution of charged states~\cite{kosovan2019},
with strong finite-size effects when the concentration of one or more reacting
particles is low, e.g., at $\tn{pH}\approx7$.

This limitation of the RxMC method has recently been addressed in the
grand-reaction method~\cite{landsgesell2020}.  By including additional
charge-neutral reactions, such as
$\emptyset \rightleftharpoons \tn{H}^+ + \tn{S}^-$,
$\tn{A} \rightleftharpoons \tn{A}^- + \tn{S}^+$, the system is able to reach
the correct equilibrium charge distribution. However, these additional
reactions increase the implementation complexity as well as the computational
cost of the method. There are now eight possible reactions in total, but
depending on the system conditions, only a fraction of these reactions
effectively sample the ionization states of the system. For example, at
$\tn{pH}\approx7$ a representative simulation volume $V < 10^7\, \tn{nm}^3$ on
average contains less than one $\tn{H}^+$ and $\tn{OH}^+$. Thus, most MC moves
involving $\tn{H}^+$ and $ \tn{OH}^-$ are rejected and only three out of the
eight possible reactions are effective.

\subsection{CR-MC method}
\label{sec:algorithm}

The premise of our new method is that it is more efficient to consider
generalized reactions in which all like-charged monovalent ions in solution
are grouped into a single particle type. These groupings keep the number of
required reactions low and thereby improve the MC sampling efficiency of
CR\@. Thus, we implement CR with three general reactions,
\begin{eqnarray}
\tn{A} &\rightleftharpoons& \tn{A}^- + \tn{X}^+\;,
\label{eq:acid_new} \\
\tn{B} &\rightleftharpoons& \tn{B}^++\tn{X}^-\;,
\label{eq:base_new} \\
\emptyset &\rightleftharpoons& \tn{X}^++\tn{X}^-\;,
\label{eq:salt_new}
\end{eqnarray}
where the $\tn{X}^\pm$ denote monovalent free ions. In the acid dissociation
reaction, Eq.~(\ref{eq:acid_new}), an acid group~(A) is ionized by
transferring a proton from the system to the reservoir, while simultaneously
transferring an ion~X$^+$ from the reservoir to the system.  If we choose X$^+$
as the dissociated cation ($\tn{X}^+={\tn{H}^+}$) and X$^-$ as the dissociated
anion ($\tn{X}^-={\tn{OH}^-}$) this scheme exactly reduces to the RxMC
method~\cite{smith1994,johnson1994}, Eqs.\ \eqref{eq:acid}--\eqref{eq:H2O}.
The scheme can be applied multiple times, separately to dissociated and salt
ions within the same simulation, in which case it becomes equivalent to the
recent grand-reaction ensemble method~\cite{landsgesell2020} discussed above.

Most coarse-grained electrolyte models, such as the restricted primitive
model~\cite{valleau80,luijten02a}, already routinely use the same interaction
potentials for all monovalent ions, so that the grouping of all monovalent
cations and protons into a single ion type, i.e.,
$\tn{X}^+=\{{\tn{H}^+},\, {\tn{S}^+}\}$, and likewise for all monovalent
anions and hydroxyl groups, $\tn{X}^-=\{{\tn{OH}^-},\, {\tn{S}^-}\}$, is
natural (Fig.~\ref{fig:model}).  The grouping operation strictly preserves the
correct grand-canonical distribution of charged states and does not affect any
equilibrium properties of the system (Appendix~\ref{sec:deriveXi}), but
reduces the number of necessary reactions. For example, Eqs.\ \eqref{eq:H2O}
and~\eqref{eq:salt}, along with the ``mixed'' reactions
$\emptyset \rightleftharpoons \tn{H}^++\tn{S}^-$ and
$\emptyset \rightleftharpoons \tn{OH}^-+\tn{S}^+$, are combined into a single
MC step, Eq.~\eqref{eq:salt_new}.  This grouping requires the interaction
potential of all ions in the group to be the same.  Thus, the grouping
operation is not applicable in systems where differences in the short-range
interaction of different ion types, such as Hofmeister-series effects, are
important. The grouping also cannot be performed between ions of different
valencies, but multivalent ions of a given valency can be grouped
straightforwardly.

The chemical potentials $\muxp$ and~$\muxm$ of the combined cation and the
combined anion species, respectively, are determined through a transformation
of the grand partition function (Appendix~\ref{sec:deriveXi}),
\begin{eqnarray}
  \label{eq:mux1}
  e^{\beta \muxp} &=& e^{\beta \mu_{\tn{H}}} + e^{\beta \mu_{\tn{S}^+}} \;, \\
  \label{eq:mux2}
  e^{\beta \muxm} &=& e^{\beta \mu_{\tn{OH}}} + e^{\beta \mu_{\tn{S}^-}} \;,
\end{eqnarray}
with $\beta =1/(k_{\tn{B}}T)$, $k_{\tn{B}}$ the Boltzmann constant, and
$\mu_{\tn{H}}$, $\mu_{\tn{S}^+}$, $\mu_{\tn{OH}}$, and $\mu_{\tn{S}^-}$ the
chemical potentials of the respective ionic species in the reservoir.
Moreover, even if the (short-range) interaction potentials of $\tn{H}^+$ and
$\tn{S}^+$ (or $\tn{OH}^-$ and $\tn{S}^-$) differ, Eqs.\ \eqref{eq:mux1}
and~\eqref{eq:mux2} are obtained in the dilute electrolyte limit where the
details of the short-range ion--ion interaction are immaterial.

The CR-MC method, Eqs.\ (\ref{eq:acid_new})--(\ref{eq:salt_new}), can also be
applied simultaneously to association or adsorption of other ions, such as
$\tn{Na}^+$, in which case $\tn{B}$ would refer to an empty surface site and
$\tn{B}^+$ would be a site occupied by~$\tn{Na}^{+}$.

\subsection{Monte Carlo algorithm}
\label{sec:MC}

To implement the scheme described, we derive the MC acceptance rate for the
general CR reactions, Eqs.~(\ref{eq:acid_new})--(\ref{eq:salt_new}), within
the framework of a grand-canonical ensemble.  There are six possible MC moves,
namely the forward and reverse move for each of the three CR reactions.
Forward and reverse moves are proposed with equal probability, leading to the
detailed balance condition
\begin{equation}
  p_{\tn{o}} p^{\tn{acc}}_{\tn{o} \to\tn{n}} =
  p_{\tn{n}} p^{\tn{acc}}_{\tn{n} \to \tn{o}} \;,
\label{eq:accrat}
\end{equation}
where $p_{\tn{o}}$ ($p_{\tn{n}}$) is the equilibrium probability of the old
(new) state and $p^{\tn{acc}}_{\tn{o} \to \tn{n}}$
($p^{\tn{acc}}_{\tn{n} \to \tn{o}}$) the corresponding acceptance rate.
Particles that participate in the reaction are chosen uniformly at random from
all eligible particles in the system.  If no suitable particles are available,
a move is rejected automatically. For clarity, in the following we use the
language of acid--base ionization equilibria, but the algorithm is general to
any ionization process.

We assume that the simulated system contains a fixed number of $\na$ acidic
groups and $\nb$ basic groups with corresponding dissociation free energies
$\Delta G_{\tn{a}}$ and $\Delta G_{\tn{b}}$, respectively. The number of free
cations $n_{\tn{X}^+}$ and free anions $\nxm$ is allowed to fluctuate via the
exchange of ions with the reservoir, which sets the temperature, the pH, and
the chemical potentials of combined ions, $\muxp$ and~$\muxm$.  The
equilibrium probability of the system being in a state with potential
energy~$E$, $\na^-$ (negatively charged) dissociated acid groups, and $\nb^+$
(positively charged) base groups, along with $\nxm$ free anions and $\nxp$
cations, is
\begin{widetext}
\begin{eqnarray}
  p(\nam,\nbp,\nxp,\nxm,E) &=& \frac{\na!}{\nam! (\na-\nam)!} e^{-\beta( \Delta G_{\tn a} +  \mu_{\tn{H}})\nam}
                               \frac{\nb!}{\nbp! (\nb-\nbp)!} e^{-\beta ({\Delta G_{\tn b} + \mu_{\tn{OH}}})\nbp} \nonumber \\
                           &\times& \frac{\left[\rho_0 \Na V e^{\beta \muxp} \right]^{\nxp}}{\nxp!}  
                                    \ \frac{\left[\rho_0 \Na V \, e^{\beta \muxm}\right]^{\nxm}}{\nxm!} \, \frac{e^{-\beta E}}{\Xi} \;.  
\label{eq:QQ}
\end{eqnarray}
The first four factors on the right-hand side capture the ionization and
combinatorial entropy of acid and base groups, where $\mu_{\tn{H}}$ and
$\mu_{\tn{OH}}$ determine the chemical potentials of dissociated cations
($\tn{H}^+$) and anions ($\tn{OH}^-$). The fifth and sixth factors represent
the ideal partition functions of the free ions, where $\rho_0$ is the
reference concentration, usually set to $\rho_0=1\,$M, $\Na$ Avogadro's
number, and $V$ the system volume. $\Xi$ is the normalizing factor, i.e., the
grand partition function of the system (see Appendix~\ref{sec:deriveXi}).

The MC acceptance rates are obtained by inserting Eq.~\eqref{eq:QQ} into
Eq.~\eqref{eq:accrat} where the ``old'' and ``new'' states correspond to the
chosen reactions, Eqs.~(\ref{eq:acid_new})--(\ref{eq:salt_new}). To simplify
the resulting expressions, we use a base-10 logarithmic representation for all
chemical potentials and dissociation constants,
$\pI_{\tn{X}^{\pm}} = -\beta\mu_{\tn{X}^{\pm}}\log_{10}e$,
$\pI_{\tn{S}^{\pm}} = -\beta\mu_{\tn{S}^{\pm}}\log_{10}e$,
$\pH = -\beta \mu_{\tn{H}}\log_{10}e$, and
$\pOH = -\beta \mu_{\tn{OH}}\log_{10}e$, with $e$ Euler's number. The chemical
potential of the combined type in the reservoir is determined by the pH of the
reservoir and $\pI_{\tn{S}^{\pm}}$ via
$10^{-\pIxp} = 10^{-\pH} + 10^{-\pI_{\tn{S}^+}}$ [see
Appendix~\ref{sec:deriveXi} and Eq.~\eqref{eq:mux}]. Likewise, the
dissociation constants are $\pKa = \beta \Delta G_{\tn a}\log_{10}e$ and
$\pKb = \beta \Delta G_{\tn b}\log_{10}e$.

The forward acid reaction consists of three steps. An acid group becomes
negatively charged, a dissociated ion ($\tn{H}^+$) is placed into the
reservoir, and an ion $\tn{X}^+$ is taken from the reservoir and placed into
the system.  The corresponding acceptance ratio is
\begin{equation}
  \frac{p^{\tn{acc}}_{(\nam)\to (\nam+1)}}{p^{\tn{acc}}_{(\nam+1) \to (\nam)}}
  = \frac{(\na - \nam) \rho_0 \Na V} {(\nam+1) (\nxp+1)} 10^{\pH-\pKa-\pIxp}
  e^{-\beta \Delta E} \;,
\label{eq:acidacc}
\end{equation}
with $\Delta E = E_{\tn{n}} - E_{\tn{o}}$ the difference in the potential
energy of the new and the old state of the system.  Thus, applying the
Metropolis scheme, we obtain the acceptance probability for the forward acid
reaction,
\begin{equation}
  p^{\tn{acc}}_{(\nam)\to (\nam+1)} = \min\left[1,\frac{(\na - \nam) \rho_0
      \Na V} {(\nam+1) (\nxp+1)} 10^{\pH-\pKa-\pIxp} e^{-\beta \Delta E}
  \right] \;.
\label{eq:MCmetro}
\end{equation}

Similarly, in the backward acid reaction an ion $\tn{X}^+$ is placed into the
reservoir, an ion $\tn{H}^+$ is taken from the reservoir and placed into the
system, and finally an acid group is neutralized. The corresponding acceptance
probability is
\begin{equation}
  p^{\tn{acc}}_{(\nam) \rightarrow (\nam-1)} = \min\left[1,\frac
    {\nam \nxp} {(\na - \nam+1) \rho_0 \Na V}10^{-\pH+\pKa+\pIxp} e^{-\beta
      \Delta E} \right] \;.
\end{equation}
The forward/backward base reaction acceptance probabilities are
\begin{equation}
  p^{\tn{acc}}_{(\nbp) \rightarrow (\nbp+1)}=\min \left[1,
    \frac{\left(\nb-\nbp\right) \rho_{0} \Na
      V}{\left(\nbp+1\right)\left(\nxm+1\right)} 10^{\pOH-\pKb-\pIxm}
    e^{-\beta \Delta E}\right]\;
\end{equation}
and
\begin{equation}
  p^{\tn{acc}}_{(\nbp) \rightarrow (\nbp-1)}=\min \left[1,
    \frac{\nbp \nxm} {\left(\nb-\nbp+1\right) \rho_{0} \Na V}
    10^{-\pOH+\pKb+\pIxm}  e^{-\beta \Delta E}\right]\;,
\end{equation}
whereas the ion pair insertion/deletion probabilities are given by
\begin{equation}
  p^{\tn{acc}}_{(n_{\tn{X}^{\pm}}) \rightarrow (n_{\tn{X}^{\pm}}+1)}=\min
  \left[1, \frac{\left(\rho_{0} \Na V\right)^2
    }{\left(\nxp+1\right)\left(\nxm+1\right)} 10^{-\pIxp-\pIxm} e^{-\beta
      \Delta E}\right]\;
\label{eq:MCsaltf}
\end{equation}
and
\begin{equation}
  p^{\tn{acc}}_{(n_{\tn{X}^{\pm}}) \rightarrow (n_{\tn{X}^{\pm}}-1)}=\min
  \left[1, \frac {\nxp \nxm} {\left(\rho_{0} \Na V\right)^2 } 10^{\pIxp+\pIxm}
    e^{-\beta \Delta E}\right]\;.
\label{eq:MCmetro2}
\end{equation}

\end{widetext}

\subsection{Performance analysis}
\label{sec:accuracy-efficiency}

The computational cost of the CR-MC method is dominated by the evaluation of
the energy~$E$.  When used with a long-range electrostatic solver, such as the
particle--particle particle--mesh (PPPM) algorithm~\cite{hockney-eastwood},
each attempted MC moves requires one evaluation of the full system energy,
provided that the energy of the original configuration was retained after a
prior MC move or MD step. Likewise, each MD time step also requires a full
calculation of long-range electrostatics, so that the computational cost of an
MC move is comparable to that of a single MD step.  The number of MC
moves~$n_{\tn{MC}}$ to be performed for every $n_{\tn{MD}}$ MD time steps
depends on the simulation setup. If the objective is to sample equilibrium
properties the main consideration is convergence to equilibrium and rapid
decorrelation of configurations. The optimal ratio is determined by the
decorrelation time scale of the system, which can be limited by either the MD
or the MC aspect of the simulation, depending on the model parameters.
However, a reasonable rule of thumb is to use $n_{\tn{MC}} \sim n_{\tn{MD}}$,
thus ensuring comparable computational cost of the MC and MD components of the
simulation. This choice guarantees that in all situations the number of
evaluations of the long-range electrostatic energy is at most twice the
optimal number and thereby yields an overall performance that is always a
large fraction of the optimal performance.
On the other hand, if the goal is to sample the correct dynamics of the
dissociation/association process with specific dissociation rates,
$n_{\tn{MC}}$ is set by these rates. For example, to model an acid with a
given dissociation rate $r_{\tn{d}}$, the ratio of acid dissociation
attempts~$n^{\tn{A}}_{\tn{MC}}$ to MD steps should be on average
$n^{\tn{A}}_{\tn{MC}} / n_{\tn{MD}} \approx \na r_{\tn{d}} \delta t$, with
$\delta t$ the MD time step.

The evaluation of the system energy is dominated by the evaluation of the
electrostatic energy. With a fast PPPM solver the computational complexity of
each MD time step and each MC move scales as $\mathcal{O}(N \log N)$, with $N$
the number of charges in the system. Moreover, the total number of MC steps
required to relax the charge distribution in the system scales approximately
linearly with the number of ionizable sites~$M$ in the system. The total
computational complexity of MC sampling is thus $\mathcal{O}(MN \log N)$,
which dominates over MD for large~$M$.  Therefore, an efficient MC algorithm
is crucial to achieve an acceptable performance when simulating large
systems. Our algorithm and implementation (Appendix~\ref{sec:lammps}) make it
possible to simulate CR phenomena in large systems containing tens of
thousands of weak acid sites~\cite{curk2021}.

To obtain a representative performance comparison between different methods,
we simulate a simple weak electrolyte consisting of $\na=500$ individual
acidic groups ($\pKa=6.5$) immersed in an implicit solvent with a dielectric
constant $\varepsilon$ within a periodic cubic box of size $L= 50l_{\tn{B}}$
with $l_{\tn{B}} = q_0^2 / (4 \pi \varepsilon \varepsilon_0 k_{\tn{B}}T)$ the
Bjerrum length, $q_0$ the elementary charge, and $\varepsilon_0$ the vacuum
permittivity.  We consider an aqueous solution ($l_{\tn{B}}=0.72\,\tn{nm}$)
with typical pH and salinity values, $\pH=7$ and $\pI_{\tn{S}^{\pm}}=2$,
corresponding to a symmetric
($\pI_{\tn{S}^{\pm}} = \pI_{\tn{S}^{+}} = \pI_{\tn{S}^{-}}$) monovalent salt
concentration $c\approx 0.01$~M.  Since we aim to evaluate the performance of
the CR algorithm, MD integration is not performed.  Electrostatic interactions
are taken into account via the PPPM algorithm with a relative force accuracy
of $10^{-5}$ and real-space cutoff $r_{\text{cut}}=10l_{\tn{B}}$.

\begin{figure}
\centering
\includegraphics[width=\narrowfigurewidth]{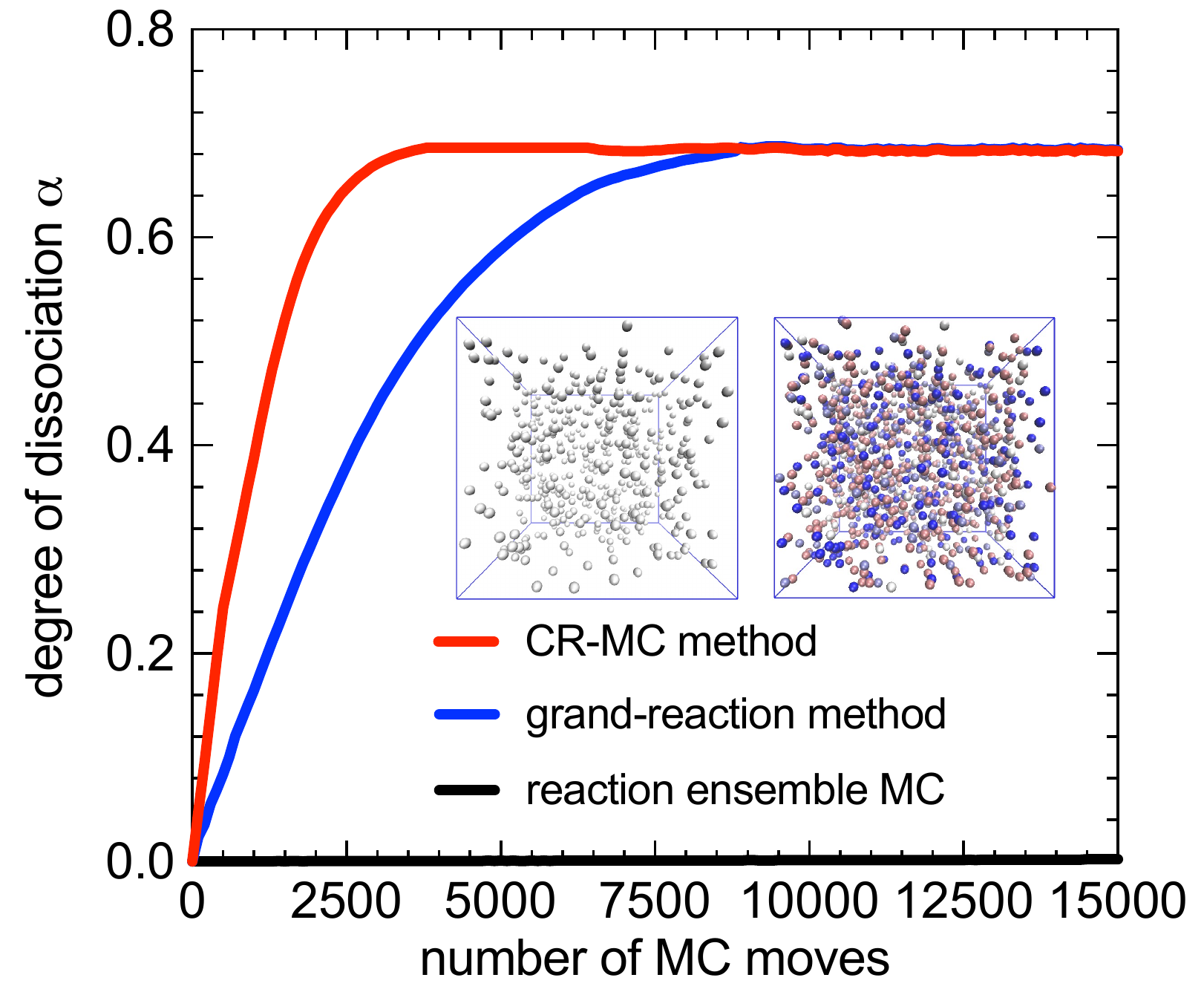}
\caption{Convergence of the average degree of dissociation~$\alpha$ obtained
  using three methods. The CR-MC method with ion grouping (red curve), the
  grand-reaction ensemble method (blue curve), and the conventional
  reaction-ensemble method (black curve), all of which can be realized using
  our LAMMPS implementation.  The inset snapshot shows the simulated system of
  a simple weak acid electrolyte (see text for details).  The CR-MC method
  results in the same thermodynamic equilibrium as the grand-reaction method,
  but significantly accelerates the convergence, whereas the reaction-ensemble
  method suffers from prohibitively strong finite-size effects and is unable
  to reach thermodynamic equilibrium.  The data is averaged over 1000
  independent runs. Parameters: $\pH=7$, $\pI_{\tn{S}^{\pm}}=2$, $\pKa=6.5$,
  $L=50l_{\tn{B}}$.}
\label{fig:efficiency}
\end{figure}

We compare the efficiency of our CR-MC method with the conventional
RxMC~\cite{smith1994,johnson1994,turner2008} and the grand-reaction ensemble
method~\cite{landsgesell2020} (Fig.~\ref{fig:efficiency}).  All three methods
differ only in the types of MC moves that are used and are simulated with our
LAMMPS implementation (see Appendix~\ref{sec:lammps}). The three methods could
also be simulated using the ESPResSo package which implements the RxMC and
grand-reaction ensemble methods. The CR-MC method can be simulated in this
package by defining rescaled chemical potentials [Eqs.\ \eqref{eq:mux1}
and~\eqref{eq:mux2}] and equilibrium constants that implement the generalized
reactions [Eqs.\ \eqref{eq:acid_new}--\eqref{eq:salt_new}], i.e., the acid
equilibrium constant $K_{\tn A} = 10^{\pH-\pKa-\pI_{\tn X^+}}$, the base
equilibrium constant $K_{\tn B} = 10^{\pOH-\pKb-\pI_{\tn X^-}}$, and the
electrolyte dissociation constant
$K_{\tn{ion}} = 10^{-\pI_{\tn X^-}-\pI_{\tn X^+}}$.

We examine the convergence of the degree of
dissociation~$\alpha = \langle \nam \rangle / \na$ averaged over 1000
independent initial configurations that are generated by randomly placing
individual neutral acid groups in the system while not allowing any overlaps,
i.e., ensuring an inter-group distance $r_{ij} > \sigma 2^{1/6} $, for all
particles in the system (Fig.~\ref{fig:efficiency} inset). The same random
configurations are used with all three methods and initially no salt ions are
present.  The conventional RxMC does not result in the correct charge
distribution due to severe finite-size effects~\cite{kosovan2019}
(Fig.~\ref{fig:efficiency}, black curve).  This issue is corrected by
completing the set of possible MC moves, i.e., by employing the grand-reaction
ensemble method~\cite{landsgesell2020} (Fig.~\ref{fig:efficiency}, blue
curve).  Lastly, the CR-MC method (Fig.~\ref{fig:efficiency}, red curve)
converges to the same degree of dissociation, but approximately three times
faster.  In this comparison, all possible MC moves are attempted with the same
probability.  Since each MC step requires the evaluation of the total
electrostatic energy---unless one or more reacting particles do not exist and
the move is immediately rejected---the computational cost of a single MC step
is expected to be very similar for all three methods. Indeed, we find that the
average CPU time per MC step is the same for the CR-MC method and the
grand-reaction ensemble method.  Instead, the origin of the slower performance
of the grand-reaction ensemble lies in the low concentration of H$^+$, on
average less than one H$^+$ ion in the simulation box, causing most MC moves
involving H$^+$ to be rejected.  Whereas this could be partially compensated
by tuning the relative frequency of the different MC moves, we note that the
grouping of all free monovalent ions of the same sign, as employed in CR-MC,
is natural in coarse-grained electrolyte models, where these ions already
routinely use the same interaction potentials.

The combined ions can at any time be ungrouped into their constituent
subspecies.  The probability~$p_i$ that any ion belongs to a a given
subspecies~$i$ is $p_i = \exp(\beta \mu_i)/\exp(\beta \mu_{\tn{X}})$. For
example, for the combined cation $\tn{X}^{+} = \{\tn{H}^+, \tn{S}^+\}$ a
fraction $p_{\tn{S}^+}=10^{-\pI_{\tn{S}^{+}} + \pIxp}$ of the $\tn{X}^{+}$
ions present in the simulation will represent salt cations~$\tn{S}^{+}$ and a
fraction $p_{\tn{H}^+} = 10^{-\pH + \pIxp}$ will represent $\tn{H}^+$
ions. The grouping of anions follows an equivalent approach.  The grouping and
ungrouping operation are exact in thermodynamic equilibrium.  When simulating
dynamics of non-equilibrium processes, the evolution of the system depends on
the transport properties of individual ions and thus the grouping operation
can no longer be applied indiscriminately. The grouping can still be applied
in non-equilibrium situations if the configurational changes of simulated
entities, e.g., polymers or nanoparticle aggregates, are slow compared to the
relaxation of the ion density distribution, such that the ion distribution can
be considered to be in quasi-equilibrium.

\section{Applications and discussion}
\label{sec:results}

\subsection{Configurations of a hydrophobic weak polyelectrolyte}
\label{subsec:hydrophobic_PE}

To demonstrate a practical application of the CR-MC method, we investigate the
equilibrium configurations of a single hydrophobic polyelectrolyte (PE) chain
consisting of $N_{\tn{m}}=80$ weak acid groups. All particles are modeled as
spheres of diameter $\sigma=l_{\tn{B}}=0.72\,\tn{nm}$ that interact via a
shifted-truncated Lennard-Jones (LJ) potential
\begin{equation}
U_{\tn{LJ}}(r_{ij}) = 
\begin{cases}
4\varepsilon_{\mathrm{LJ}}\left[
  \left(\frac{\sigma}{r_{ij}}\right)^{12}
  -\left(\frac{\sigma}{r_{ij}}\right)^{6}+C
\right] & \; r_{ij}\leq r_{\tn{cut}} \\
 0      & \; r_{ij} >   r_{\tn{cut}} \;.
\end{cases}
\label{eq:LJ}
\end{equation}
We use the standard cutoff $r_{\tn{cut}}=2.5\sigma$ and shift
$C=2.5^{-6}-2.5^{-12}$ for intra-PE interactions to simulate hydrophobic
effects, while all other short-range interactions are purely repulsive, namely
$r_{\tn{cut}}=2^{1/6}\sigma$ and $C=1/4$.  Table~\ref{tab:weak-PE} summarizes
the LJ interaction parameters, where the Lorentz--Berthelot mixing rule is
used.

\begin{table}[b]
  \caption{Lennard-Jones parameters $\varepsilon_{\mathrm{LJ}}$,
    $r_{\tn{cut}}$ [Eq.~\eqref{eq:LJ}] for the hydrophobic PE system. We set
    $\sigma=l_{\tn{B}}$ and $\epsilon_{\tn{ii}}=k_{\tn{B}}T$. The subscripts
    ``i'' and ``m'' refer to ions and monomers, respectively.}
\label{tab:weak-PE}
\begin{tabular}{c|c|c}
 \hline type & acid monomer & free ion \\
\hline acid monomer & $\varepsilon_{\tn{mm}}$, $2.5\sigma$ & $\sqrt{\varepsilon_{\tn{mm}}\varepsilon_{\tn{ii}}}$, $2^{1/6}\sigma$\\
\hline free ion & $\sqrt{\varepsilon_{\tn{mm}}\varepsilon_{\tn{ii}}}$, $2^{1/6}\sigma$ & $\varepsilon_{\tn{ii}}$, $2^{1/6}\sigma$   \\
\hline
\end{tabular}
\end{table}

Neighboring monomers along the chain are bonded through a harmonic potential,
\begin{equation}
  U_{\tn{bond}}(r_{ij})= K({r_{ij}}-R_0)^2\;,
\label{eq:bond}
\end{equation}
with spring constant $K=400 k_{\tn{B}}T/{\sigma}^2$ and bond length
$R_0=2^{1/6}\sigma$. The relatively large spring constant ensures that the
intra-chain electrostatic repulsion cannot noticeably affect the contour
length of the chain. The PE is simulated in a periodic cubic box with
dimensions $L=100l_{\tn{B}}$. Electrostatic interactions are treated in the
same manner as in Sec.~\ref{sec:accuracy-efficiency}.  The temperature is
controlled by a Langevin thermostat with damping time
$\tau=[m\sigma^2/(k_{\tn{B}}T)]^{1/2}$, where $m$ is the ion mass. The
positions and velocities are updated using the velocity-Verlet algorithm with
a time step of $\delta t = 0.005\tau$. After every $n_{\tn{MD}}=400$ MD steps,
we perform $n_{\tn{MC}}=200$ MC steps.  We start from a random configuration
and equilibrate the system for $10^3\tau$. The subsequent production runs last
for $2\times10^5\tau$, during which the configurations are sampled every
$20\tau$.

\begin{figure*}
\centering 
\includegraphics[width=\figurewidth]{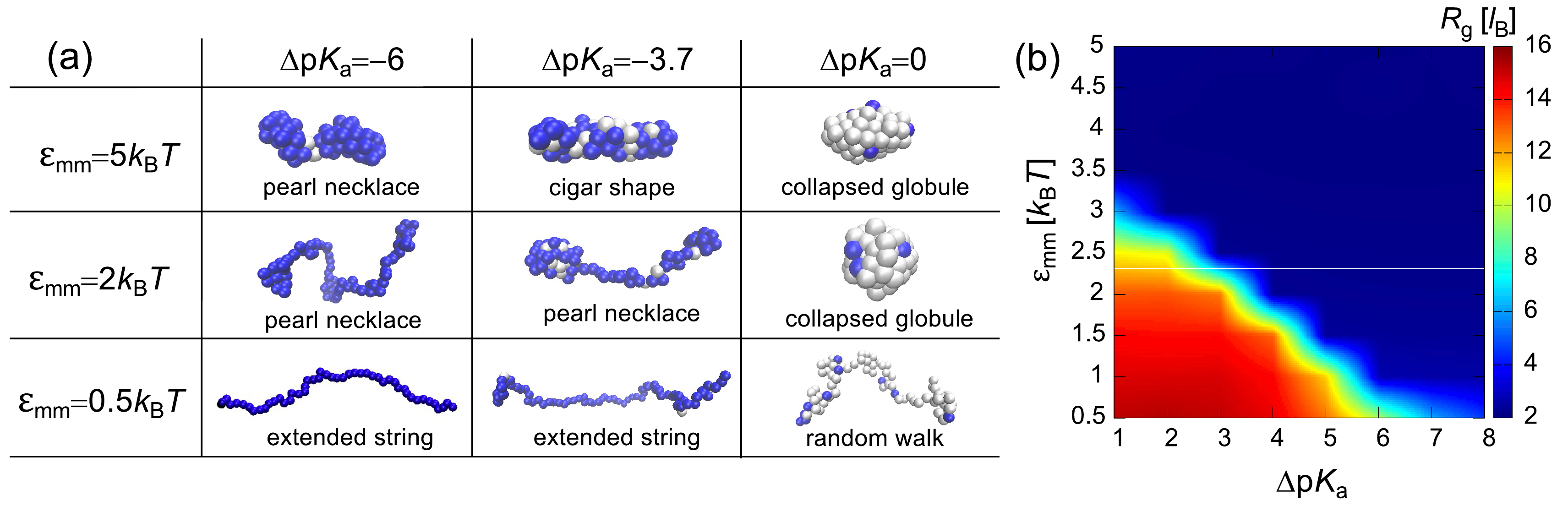}
\caption{Equilibrium conformations of a hydrophobic weak polyelectrolyte.
  (a)~Representative configurations at different dissociation strength
  $\Delta\pKa=\pKa-\tn{pH}$ and short-range (hydrophobic)
  attraction~$\epsilon_{\tn{mm}}$. Charged bead are shown in blue and neutral
  beads in white.  Free ions are not shown for better visualization of the PE
  conformations. (b)~The corresponding radius of gyration~$R_{\text{g}}$ shows
  an abrupt transition between extended (large $R_{\tn{g}}$, red shading) and
  collapsed (low $R_{\tn{g}}$, blue shading) structures. In all cases we used
  $\pH=7$ and $\pI_{\tn{S}^{\pm}}=3$.}
\label{fig:weak-PE1}
\end{figure*}

The competition between the electrostatic monomer repulsion, which promotes
polymer expansion, and the short-range monomer attraction representing
hydrophobicity, which promotes collapse, leads to a rich conformational
behavior of a hydrophobic PE chain.  Indeed, by tuning the values of
$\Delta\pKa=\pKa-\text{pH}$ and the monomer
attraction~$\varepsilon_{\tn{mm}}$ we reproduce the five types of structures
reported in earlier simulation work~\cite{ulrich2005a}, namely the random walk,
extended string, collapsed globule, ``cigar shape,'' and pearl-necklace
conformations (Fig.~\ref{fig:weak-PE1}a).

In the case of strong dissociation, $\Delta\pKa=-6$, the PE chain is nearly
fully charged and exhibits the well-known conformations of strong PEs, from a
pearl-necklace structure at strong attraction (large $\varepsilon_{\tn{mm}}$)
to an extended string-like structure at weaker attraction
(Fig.~\ref{fig:weak-PE1}a, left column). At intermediate degrees of
dissociation, $\Delta\pKa=-3.7$, the polyelectrolyte is only partially
charged, leading to weaker electrostatic repulsion and consequently more
compact structures, such as a ``cigar''-shaped structure
(Fig.~\ref{fig:weak-PE1}a, middle column).  Furthermore, weak dissociation,
$\Delta\pKa=0$, leads to a largely uncharged PE\@. Under these conditions, the
chain collapses into a globule at strong monomer attraction, but behaves as a
neutral, random coil at weak attraction (Fig.~\ref{fig:weak-PE1}a, right
column).

\begin{figure}[b]
\centering 
\includegraphics[width=\narrowfigurewidth]{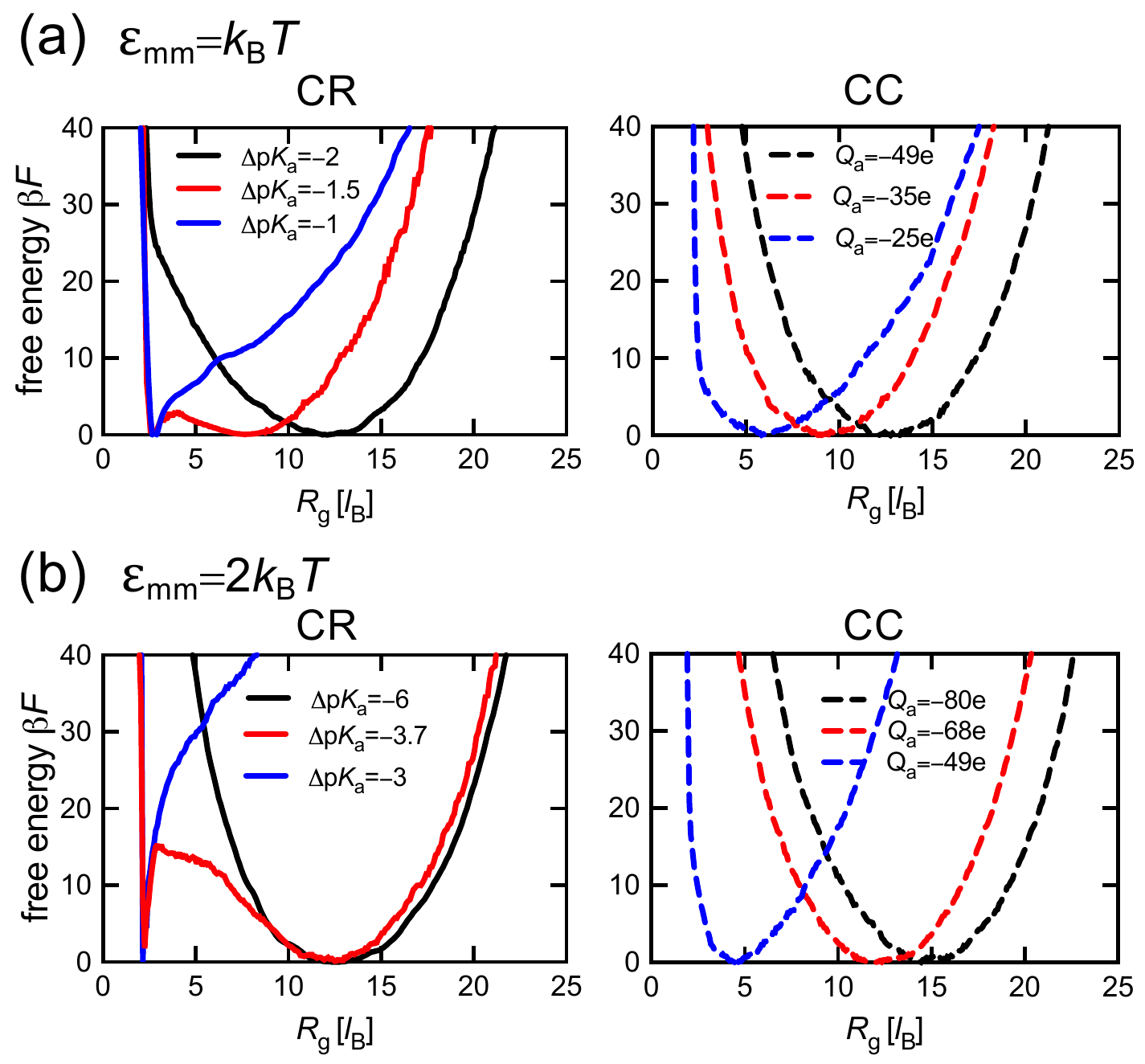}
\caption{Free-energy profiles as a function of $R_{\tn{g}}$ for a hydrophobic
  weak polyelectrolyte exhibit two stable minima, which indicates the
  coil--globule transition detected in Fig.~\ref{fig:weak-PE1}(b) is
  discontinuous (solid lines), while the corresponding constant-charge (CC)
  approximation (dashed lines) results in a single minimum. In (a) and (b), we
  set $\epsilon_{\tn{mm}}=1.0k_{\tn{B}}T$ and
  $\epsilon_{\tn{mm}}=2.0k_{\tn{B}}T$, respectively.}
\label{fig:weak-PE2}
\end{figure}

We note that the average radius of gyration~$R_{\text{g}}$ characterizing
these conformations displays an abrupt change in the parameter space span by
$\varepsilon_{\tn{mm}}$ and $\Delta\pKa$ (Fig.~\ref{fig:weak-PE1}b).  As our
CR-MC method is implemented in LAMMPS, it can be directly combined with
standard free-energy methods. Thus, we examine the nature of the variation in
shape by calculating the free-energy profiles as a function of
$R_{\tn{g}}$. We apply the metadynamics method~\cite{laio2002} implemented in
the COLVARS library~\cite{colvars2013} of LAMMPS, using a bin size
$0.05\sigma$ and a “Gaussian-hill" weight $0.005k_{\tn{B}}T$, with hills
deposited every $10$ MD time steps. We find that under CR conditions the
free-energy profile exhibits two minima at the transition point
(Fig.~\ref{fig:weak-PE2}), indicating a coil--globule transition that is first
order (with the caveat that a true thermodynamic transition would require the
limit of infinite chain length). Prior theoretical~\cite{raphael1990} and
simulation~\cite{uyaver2003} investigations indeed suggested that CR effects
could lead to a first-order coil--globule transition of weak PEs, but to our
knowledge this is the first time that this has been confirmed by a free-energy
calculation.  Repeating the simulations in the corresponding CC approximation,
where each chain bead has a fixed charge~$Q_{\tn{a}}/N_{\tn{m}}$ ($Q_{\tn{a}}$
the average charge of the weak PE obtained from an equilibrium CR simulation),
we always find a single free-energy minimum.  Thus, while the observed
structures are not new, CR leads to transitions between these structures that
are qualitatively different from those observed in the conventional CC
approximation.

\subsection{Potential of mean force between charged nanoparticles}

Charge regulation effects have been shown to reduce the electrostatic
repulsion between like-charged particles~\cite{takae2018,dosSantos2019} and
increase the attraction between a large particle coated with dissociable sites
and a small ion~\cite{curk2021}.  However, the interaction between two
oppositely charged particles has not been explored in detail, even though this
arrangement provides a prototypical model for investigating CR effects on
electrostatic protein--protein
interactions~\cite{lund2005a,lund2013,roosen-runge2014}.

As an illustration, we calculate the potential of mean force (PMF) between two
oppositely charged nanoparticles of radius~$R=3.5l_{\tn{B}}$ immersed in an
implicit aqueous solvent characterized by $l_{\tn{B}}=0.72$~nm and
$\tn{pH}=7$. The solution also contains free ions (salt, protons, and hydroxyl
ions) of diameter~$\sigma=l_{\tn{B}}$.  The weak acid and weak base groups on
the surface of the nanoparticles have dissociation constants
$\pKa=\pKb=6.5$. A set of 256 acid/base groups is uniformly distributed on,
and rigidly attached to, the shell (radius $3l_{\tn{B}}$) of each
sphere~\cite{curk2021}.  We examine two salt concentrations, namely
$\pI_{\tn{S}^{\pm}} = 6$, $c\approx10^{-6}$~M, representing deionized water, and
$\pI_{\tn{S}^{\pm}} = 1$, $c\approx0.1$~M, representing a physiological saline solution.
In the former case, we utilize a system size $L=1000l_{\tn{B}}$ and in the
latter $L = 40l_{\tn{B}}$.  This ensures that the number of free ions in the
solution greatly exceeds the charge on the individual nanoparticles, thus
avoiding spurious long-range electrostatic interactions between periodic
images.  The excluded-volume interactions are modeled through the expanded LJ
potential,
\begin{equation}
U_{\tn{LJ}}(r_{ij})= 
\begin{cases}
  4\varepsilon_{\mathrm{LJ}} \left[\left(\frac{\sigma}{{r_{ij}}-\Delta}\right)^{12}-\left(\frac{\sigma}{{r_{ij}}-\Delta}\right)^{6}+\frac{1}{4}\right] & \; r_{ij}\leq r_{\tn{c}}^* \\
  0 & \; r_{ij}> r_{\tn{c}}^*\,,
\end{cases}
\label{eq:LJe}
\end{equation}
with $\Delta$ the expanded distance and $r_{\tn{c}}^*=\Delta + 2^{1/6}\sigma$
the cutoff.  The interaction parameters for the different combinations of
particle types are listed in Table~\ref{tab:PMF-LJ}.

\begin{table}[h]
  \caption{Lennard-Jones interaction parameters in the system of colloidal
    nanoparticles (radius~$R$) and free ions (radius~$r$). We always use
    $\varepsilon_{\mathrm{LJ}} = k_{\tn{B}}T$, $\sigma=2r$, and cutoff
    $r_{\text{c}}^*= \Delta + 2^{1/6}\sigma$.}
\label{tab:PMF-LJ}
\begin{tabular}{c|l|l}
\hline type & particle $(R)$ & free ions $(r)$ \\
\hline particle $(R)$ & $\Delta=2 R-2 r$ & $\Delta=R-r$ \\
\hline free ions $(r)$ & $\Delta=R-r$ &  $\Delta=0$ \\
\hline
\end{tabular}
\end{table}

The temperature~$T$ is controlled by a Langevin thermostat with damping
time~$20\tau$, where the unit time $\tau$ (Sec.~\ref{subsec:hydrophobic_PE})
is based upon the mass~$m$ of the ions and dissociable groups. The total
nanoparticle mass is $257m$.  After equilibrating the system for
$5 \times 10^3\tau$, the production runs last for $5 \times 10^5\tau$ with MD
time step $\delta t = 0.005\tau$. After every $400$ MD steps,
we perform $200$ MC steps.

\begin{figure*}
\centering 
\includegraphics[width=\figurewidth]{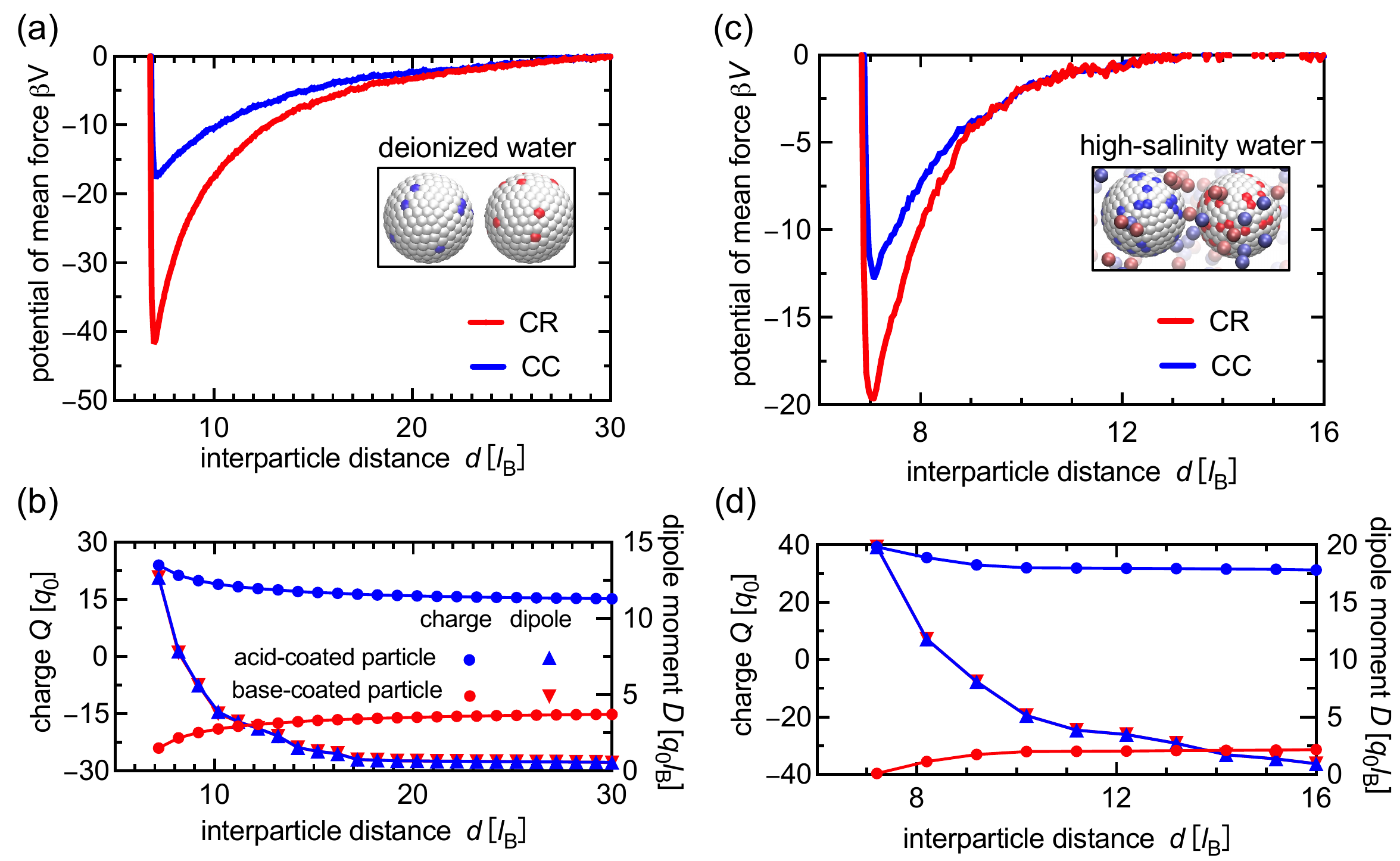}
\caption{Effect of charge regulation on the interactions between two
  oppositely charged nanoparticles, modeled as an acid-coated and a
  base-coated particle. The respective dissociation constants are $\pKa=6.5$
  and $\pKb=6.5$, while the $\pH$ is fixed at~$7$. (a)~Potential of mean force
  (PMF) between the charge regulating (CR) particles (red curve) and
  comparison to the constant charge (CC) approximation (blue curve), when the
  particles are immersed in deionized water ($\pI_{\tn{S}^{\pm}}=6$). Inset
  shows a typical configuration of an acid-coated particle displaying
  dissociated (blue) and neutral (white) acid groups, accompanied by a
  base-coated particle with positively charged (red) and neutral (white) base
  groups. In the CC simulation, the particle charges are
  $Q_{\tn{acid}} = -Q_{\tn{base}} = -14q_0$ , obtained from the equilibrium
  surface charge of isolated particles ($d\to \infty$).  (b)~Variation of the
  total charge on the acid-coated (blue circles) and the base-coated (red
  circles) CR nanoparticle with varying separation~$d$. The total dipole
  moment of both particles is shown as well (blue and red triangles,
  respectively).  (c)~Counterpart of panel~(a), showing the PMF of the
  nanoparticle pair immersed in physiological saline solution
  ($\pI_{\tn{S}^{\pm}}=1$). In this case, the particle charge in the CC case
  is $Q_{\tn{acid}}=-Q_{\tn{base}}=-29q_0$. Inset shows a typical
  configuration, now also featuring free cations (light red) and anions (light
  blue). (d)~Counterpart of panel~(b), illustrating the variation of
  nanoparticle charge with separation at $\pI_{\tn{S}^{\pm}}=1$. Panels (b)
  and~(d) demonstrate that the enhanced attraction of nanoparticles under CR
  conditions originates from the increased net charge at short separations as
  well as the nonuniform distribution of the ionized surface groups. Further
  computational details are provided in the main text.}
\label{fig:PMF-colloid}
\end{figure*}

To investigate the role of CR effects, we calculate the PMF of two colloidal
nanoparticles using the metadynamics technique described in
Sec.~\ref{subsec:hydrophobic_PE} and compare it to the PMF in the CC
approximation. The latter is realized by placing single charges
$Q_{\tn{acid}}$ and~$Q_{\tn{base}}$, obtained from independent simulations of
isolated acid-coated and base-coated nanoparticles at the above-mentioned
conditions, at their respective centers of mass. The PMF profiles
(Fig.~\ref{fig:PMF-colloid}a,c) show that CR can enhance pairwise interactions
about twofold compared to the CC situation. This enhancement is a result of
both the change in the average charge per particle and the nonuniform
surface-charge distribution characterized by the induced dipole moment
(Fig.~\ref{fig:PMF-colloid}b,d). Notably, the enhancement of pairwise
interactions by CR not only occurs in deionized water
(Fig.~\ref{fig:PMF-colloid}a,b), but persists at physiological salt
concentration ((Fig.~\ref{fig:PMF-colloid}c,d). This is markedly different
from dielectric effects, which are effectively screened under such
conditions~\cite{antila18a,yuan2020}. This example illustrates that CR effects
must be generally taken into account when modeling bio-macromolecular
interactions that typically occur at physiological salt conditions.

\section{Summary}
\label{sec:conclusions}

We have introduced the CR-MC method, a MC scheme that makes possible the
efficient and accurate calculation of charge regulation in solvated systems.
The method is most suitable for coarse-grained models with implicit solvent,
where the details of the short-range ion--ion interaction, such as
Hofmeister-series effects, can be neglected.  By grouping all like-charged
free monovalent ions into a single particle type and allowing monovalent salt
to participate in acid--base reactions, our CR-MC method outperforms previous
approaches (constant-pH method~\cite{reed1992,radak2017},
RxMC~\cite{smith1994,johnson1994,turner2008}, and grand-RxMC
method~\cite{landsgesell2020}) in applicable parameter range or efficiency.

We have implemented the CR-MC method within the LAMMPS~\cite{plimpton95} MD
package. The implementation is parallelized and compatible with existing
LAMMPS functionalities, such as rigid-body dynamics and free-energy
calculations, and thus markedly lowers the entry barrier to incorporating
charge regulation effects into MD simulations. We emphasize that this enables
\emph{self-consistent} calculations in which the instantaneous distribution of
particles and charges determines the electrostatic forces that drive the time
evolution of the system, and conversely the resulting distributions affect the
charge states of the particles. The LAMMPS implementation also supports
RxMC~\cite{smith1994,johnson1994,turner2008} and
grand-RxMC~\cite{landsgesell2020} methods.

We have demonstrated the capabilities of our approach by determining the
conformations of a hydrophobic weak polyelectrolyte for different dissociation
conditions as well as the corresponding free-energy profiles as a function of
its radius of gyration.  We found that CR effects lead to the coexistence of
two stable states at the coil--globule transition, implying a discontinuous
coil--globule transition, thus corroborating previous predictions~\cite{uyaver2003}. 
Interestingly, this discontinuous transition
vanishes in the usual CC approximation that ignores the fluctuations of
individual charges on the polyelectrolyte.  As a second example, we calculated
the PMF between an acid-coated and a base-coated colloidal nanoparticle,
demonstrating that CR effects give rise to an approximately twofold increase
in the attractive interaction at both low and high salinity. These examples
show that CR effects can markedly alter the behavior of charged systems and
demonstrate the importance of an accurate CR solver.

The CR-MC method allows the modeling of simple reactions and charge
redistribution in a broad range of coarse-grained, solvated systems, such as
polyelectrolytes, proteins, membranes, and nanoparticles. Although we have
focused on simulating acid--base ionization equilibria, the method is general
and can be used to model any two-state association/dissociation process.

\begin{acknowledgments}
  This material is based upon work supported by the E.U. Horizon 2020 program
  under the Marie Sk\l{}odowska-Curie fellowship No.~845032 and by the
  U.S. National Science Foundation through Grant No.\ DMR-1610796.  J.Y.
  acknowledges the support of a Professional Development Fellowship offered by
  Shanghai Jiao Tong University. We thank Roman Sta{\v n}o and David Beyer for
  testing our LAMMPS implementation.
\end{acknowledgments}

\section*{Data Availability}

The data that support the findings of this study are available within the
article. The source code of our CR-MC implementation is available via the
standard LAMMPS repository, see Appendix~\ref{sec:lammps}.

\appendix

\section{Derivation of the grand-canonical ensemble with ion grouping}
\label{sec:deriveXi}

We show that the proposed grouping of different ions, which is at the core of
the efficiency gain provided by the CR-MC method, within the primitive
electrolyte model preserves the correct grand-canonical distribution of
concentrations and leads to Eq.~\eqref{eq:QQ}. The grand-canonical ensemble of
states for a system with volume~$V$ and temperature~$T$ in contact with a
reservoir containing $M$ different particle types is determined by $M$
chemical potentials~$\mu_i$, denoted in vector notation as
$\boldsymbol{\mu} = [\mu_1,\cdots, \mu_{M}]$. The grand-canonical partition
function describing this ensemble,
\begin{equation}
\Xi (\boldsymbol{\mu},V,T) = \sum_{\mathbf{N}=0}^{\infty} e^{\beta \boldsymbol{\mu} \cdot \mathbf{N} }Q (\mathbf{N},V,T) \;,
\label{eq:Xi}
\end{equation} 
is obtained by summing over all $\mathbf{N}=[N_1, \cdots, N_M]$ possible
numbers of particles of each type in the system, where the summation denotes a
nested sum,
$\sum_{\mathbf{N}}[\cdot] = \sum_{N_1} \sum_{N_2} \cdots \sum_{N_M} [\cdot]$,
and $\beta = 1/(k_{\tn{B}}T)$ with $k_{\tn{B}}$ the Boltzmann constant.  The
canonical partition function,
\begin{equation}
Q (\mathbf{N},V,T) = \prod_{l=1}^{M} \frac{1}{N_l!} \left(\frac{V}{\Lambda^3}\right)^{N_l}  \mathcal{I}(\mathbf{N},V,T) \;,
\label{eq:Q}
\end{equation}
contains the product performed over the ideal-gas contributions of individual
particle types, with the reference length scale
$\Lambda = (\rho_0 N_{\tn{A}})^{-1/3}$ set by the reference concentration
$\rho_0=1\;\tn{M}$. The configurational contribution,
\begin{equation}
 \mathcal{I}(\mathbf{N},V,T) = \int d\mathbf{r}^N e^{-\beta E(\mathbf{r}^N)}\;,
  \label{eq:I}
\end{equation} 
is obtained by integrating over the positions of all $N=\sum_l N_l$ particles
in the system, with $E(\mathbf{r}^N)$ the potential energy of the system that
depends on the positions~$\mathbf{r}$ of all $N$ particles.

Within the primitive model electrolyte, all monovalent ions use the same
short-range interaction potential. Therefore, exchanging one cation type for
another cation type leaves the potential energy of the system unchanged. For
example, if type~$i$ represents $\tn{H}^+$, type~$j$ represents salt
cation~$\tn{S}^+$, and $N_i > 0$, the configuration integral is invariant under
changing of ion types,
\begin{equation}
 \mathcal{I}( [ \cdot, N_{i}, N_{j}, \cdot] ,V,T) =  \mathcal{I}( [ \cdot, N_{i}-1, N_{j}+1, \cdot], V,T) \;.
  \label{eq:Iswap}
\end{equation} 
Therefore, by induction, $\mathcal{I}$ is a function only of the sum $N_i+N_j$,
\begin{equation}
\mathcal{I} = \mathcal{I}( [\cdot,N_{i} + N_{j}, \cdot],V,T)\;.
\end{equation}
Using this property and Eq.~\eqref{eq:Q} we rewrite Eq.~\eqref{eq:Xi} as
\begin{equation}
\Xi (\boldsymbol{\mu},V,T) = \sum_{\mathbf{N'}=0}^{\infty}  e^{\beta \boldsymbol{\mu}' \cdot \mathbf{N'}} \prod_{l \ne i,j}^{M} \left[ \frac{1}{N_{l}!} \left(\frac{V}{\Lambda^3}\right)^{N_{l}} \right]  \mathcal{S} \;,
\label{eq:Xi2}
\end{equation} 
where $\boldsymbol{\mu}'$ and $\mathbf{N}'$ contain all particle types except for
$i$ and~$j$ and $\mathcal{S}$ denotes the sum over elements $i$ and~$j$,
\begin{equation}
 \mathcal{S} = \sum_{N_i=0}^{\infty} \sum_{N_{j}=0}^{\infty}   \frac{e^{\beta \mu_i N_i + \beta \mu_j N_j}}{N_i ! N_j !} \left(\frac{V}{\Lambda^3}\right)^{N_i+N_j}   \mathcal{I}\;.
\label{eq:S}
\end{equation}
This sum can be rewritten as a sum over $N_{\tn{X}} = N_i + N_j$, 
\begin{equation}
 \mathcal{S} = \sum_{N_{\tn{X}}=0}^{\infty} \sum_{N_{i}=0}^{N_{\tn{X}}}   \frac{e^{\beta \mu_i N_i + \beta \mu_j (N_{\tn{X}}-N_i)}}{N_i ! (N_{\tn{X}} - N_i) !} \left(\frac{V}{\Lambda^3}\right)^{N_{\tn{X}}}   \mathcal{I} \;.
\label{eq:S2}
\end{equation} 
Since $\mathcal{I}$ does not explicitly depend on $N_i$, the inner sum can be
recognized as a binomial expansion and Eq.~\eqref{eq:S2} can be
written as
\begin{equation}
 \mathcal{S} = \sum_{N_{\tn{X}}=0}^{\infty}  \left(e^{\beta \mu_i} + e^{\beta \mu_j}\right)^{N_{\tn{X}}} \frac{1}{N_{\tn{X}}!} \left(\frac{V}{\Lambda^3}\right)^{N_{\tn{X}}}  \mathcal{I} \;.
\label{eq:S3}
\end{equation} 
This represents the grand-canonical partition function of a combined type
$\tn{X}$ with chemical potential
\begin{equation}
\mu_{\tn{X}} = k_{\tn{B}}T \ln \left[ e^{\beta \mu_i} +  e^{\beta \mu_j}\right] \;.
\label{eq:mux}
\end{equation}
Insertion of Eq.~\eqref{eq:S3} into Eq.~\eqref{eq:Xi2} yields a reduced
partition function in $M-1$ particle types that is identical to the original
full partition function in $M$ particle types, Eq.~\eqref{eq:Xi}. Thus, the
CR-MC method, Eqs.~\eqref{eq:acidacc}--\eqref{eq:MCmetro2}, which samples the
statistical ensemble with combined ion types, Eq.~\eqref{eq:QQ}, leads to
exactly the same equilibrium observables as a Monte Carlo scheme (e.g., the
scheme of Ref.~\onlinecite{landsgesell2020}) in which all ions are treated
separately.

\section{LAMMPS implementation and usage}
\label{sec:lammps}

We have implemented the CR-MC method described in Sec.~\ref{sec:algorithm}
within the LAMMPS MD package. Our implementation is open source and
distributed under the GNU General Public License (GPL)\@. It is available from
the central LAMMPS repository (\url{https://lammps.sandia.gov/}), including
documentation and examples.

This LAMMPS implementation performs MC sampling of ionization states [Eqs.\
\eqref{eq:MCmetro}--\eqref{eq:MCmetro2}]. The only input parameters required
are the equilibrium constants ($\tn{p}K$), chemical potentials (pH, pOH) of
dissociated ions, and the chemical potential of inserted ions, $\pIxp$ and
$\pIxm$.  The implementation is general. For example, choosing $\pIxp = \pH$
and $\pIxm= \pOH$ would perform canonical sampling of standard reactions
[Eqs.~\eqref{eq:acid}--\eqref{eq:H2O}] following the RxMC approach for a
closed system.  The method can be invoked repeatedly to perform reactions with
different types of ions within a single simulation, thus enabling simulation
in the grand-reaction ensemble. To set up the CR-MC method presented in this
work in our LAMMPS implementation, the dissociated ions and salt ions are
combined into a single types of cations~$\tn{X}^{+}$ and anions~$\tn{X}^{-}$,
cf.\ Eq.~\eqref{eq:mux}.  Moreover, the implementation supports setting a
variable (i.e., time-dependent) pH of the reservoir and can thus, for example,
be used to study the response of a system to an increase in pH.

\section{Numerical validation}
\label{sec:Weak_acid_dissociation}

To confirm the correct functioning of our CR-MC implementation, we simulate
weak acid dissociation over a wide range of parameters and compare our results
to the grand-reaction ensemble approach~\cite{landsgesell2020} implemented in
the ESPResSo MD package~\cite{limbach2006} (version 4.0.2).  We examine a test
system containing $n_{\tn{A}}=100$ acid groups immersed in an aqueous solution
at room temperature. We set $L=30\lb$, resulting in a similar density of acid
groups as in Sec.~\ref{sec:accuracy-efficiency}. All other interaction and
system parameters are also described in Sec.~\ref{sec:accuracy-efficiency}.

We explore the behavior of the system at non-neutral pH values,
$\tn{pH} \ne \tn{pOH}$.  In this case, charge neutrality of the reservoir
implies that the chemical potentials of cations and anions (other than
$\tn{H}^+$ and $\tn{OH}^-$) in the reservoir are different and must be
specified separately.  We assume that a non-neutral pH is obtained using a
small monovalent acid or base, e.g., HCl or NaOH, which allows us to group the
free negatively charged acid with the other monovalent anions into a single
particle type, and likewise to group the free positively charged base with
free cations.  The chemical potential (in the $\log_{10}$ representation) of
these additional acid anions ($\pI_{\tn{A}^-}$) is related to the pH,
$10^{-\pI_{\tn{A}^-}} = 10^{-\pI_{\tn{S}^{\pm}}} +
10^{-\pH}-10^{-\pOH}$. Likewise, for basic solutions the chemical potential of
the additional base cations ($\pI_{\tn{B}+}$) is determined by
$10^{-\pI_{\tn{B}^+}} = 10^{-\pI_{\tn{S}^{\pm}}} + 10^{-\pOH}-10^{-\pH}$.
Using the grouping operation [Eq.~\eqref{eq:mux}] the chemical potential of
the combined ion type is thus determined by $\pI_{\tn{S}^{\pm}}$ and pH via
\begin{equation}
  10^{-\tn{pI}_{\tn{X}^{\pm}}} = 10^{-\pI_{\tn{S}^{\pm}}} + 10^{-\min[\tn{pH},\, \tn{pOH}]}\;,
  \label{eq:pIXgen}
\end{equation}
where the first term on the right-hand side takes into account the symmetric
monovalent salt, while the second term captures the dissociated ions as well
as any free acid/base groups or ions that must be present to maintain a
charge-neutral solution at a non-neutral pH.

\begin{figure*}[hbt]
  \centering \includegraphics[width=\figurewidth]{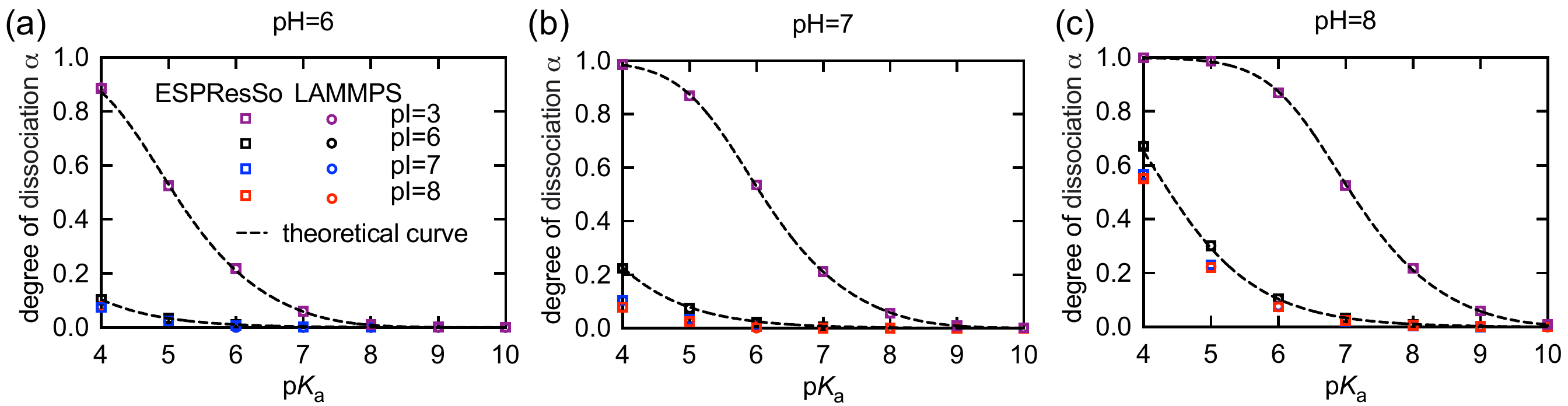}
  \caption{Average degree of dissociation~$\alpha$ of monomeric acid groups as
    a function of $\pKa$. Data are shown for different values of
    $\pI=\pI_{\tn{S}^{\pm}}$, which determines the chemical potential of
    monovalent salt, and at three different pH (panels a--c). Square symbols
    are obtained by the grand-reaction method~\cite{landsgesell2020}, which
    simulates the full set of six reactions in the ESPResSo MD
    package. Circles represent data obtained by the CR-MC method described in
    this work and implemented in LAMMPS. For all parameters considered, the
    two data sets show perfect agreement. The simulation data are further
    validated by a numerical approach that assumes ideal solution conditions
    and couples the Gibbs--Donnan equilibrium to the Henderson--Hasselbalch
    ionization equilibrium (dashed lines at $\pI=3$ and $\pI=6$).}
\label{fig:weak-acid}
\end{figure*}

In the numerical comparison, the temperature is controlled by a Langevin
thermostat with damping time~$\tau$. The positions and velocities are updated
using the velocity-Verlet algorithm with time step $\delta t =
0.01\tau$. After every $n_{\tn{MD}}=400$ MD steps we perform $n_{\tn{MC}}=200$
MC steps. We start from a random configuration and equilibrate the system for
$10^3\tau$. The subsequent production runs last for $2 \times 10^5\tau$,
during which the configuration averages are sampled every $20\tau$, yielding
the average degree of dissociation~$\alpha$ (Fig.~\ref{fig:weak-acid}).
As expected, increasing $\Delta\pKa=\pKa-\pH$ results in a lower $\alpha$,
whereas adding more salt (decreasing~$\pI_{\tn{S}^{\pm}}$) promotes acid
dissociation as the additional salt screens the electrostatic repulsion
between charged acid groups. In all cases, our implementation produces results
that are statistically identical to those obtained using the ESPResSo
package. We find our LAMMPS implementation to be about three times faster per
MC step, which we attribute primarily to the CR-MC implementation requiring a
single electrostatic energy evaluation per MC move, whereas the current
reaction ensemble implementation in ESPResSo (version 4.0.2) calls the full
energy evaluation twice per MC move.  We emphasize that this difference in
execution time per MC step is \emph{in addition} to the more rapid
decorrelation of the configurations resulting from the improved sampling of
the CR-MC method (Fig.~\ref{fig:efficiency}).  The combined effect of these
two enhancements results in an approximately 9-fold acceleration.

\begin{figure}
  \centering \includegraphics[width=0.45\textwidth]{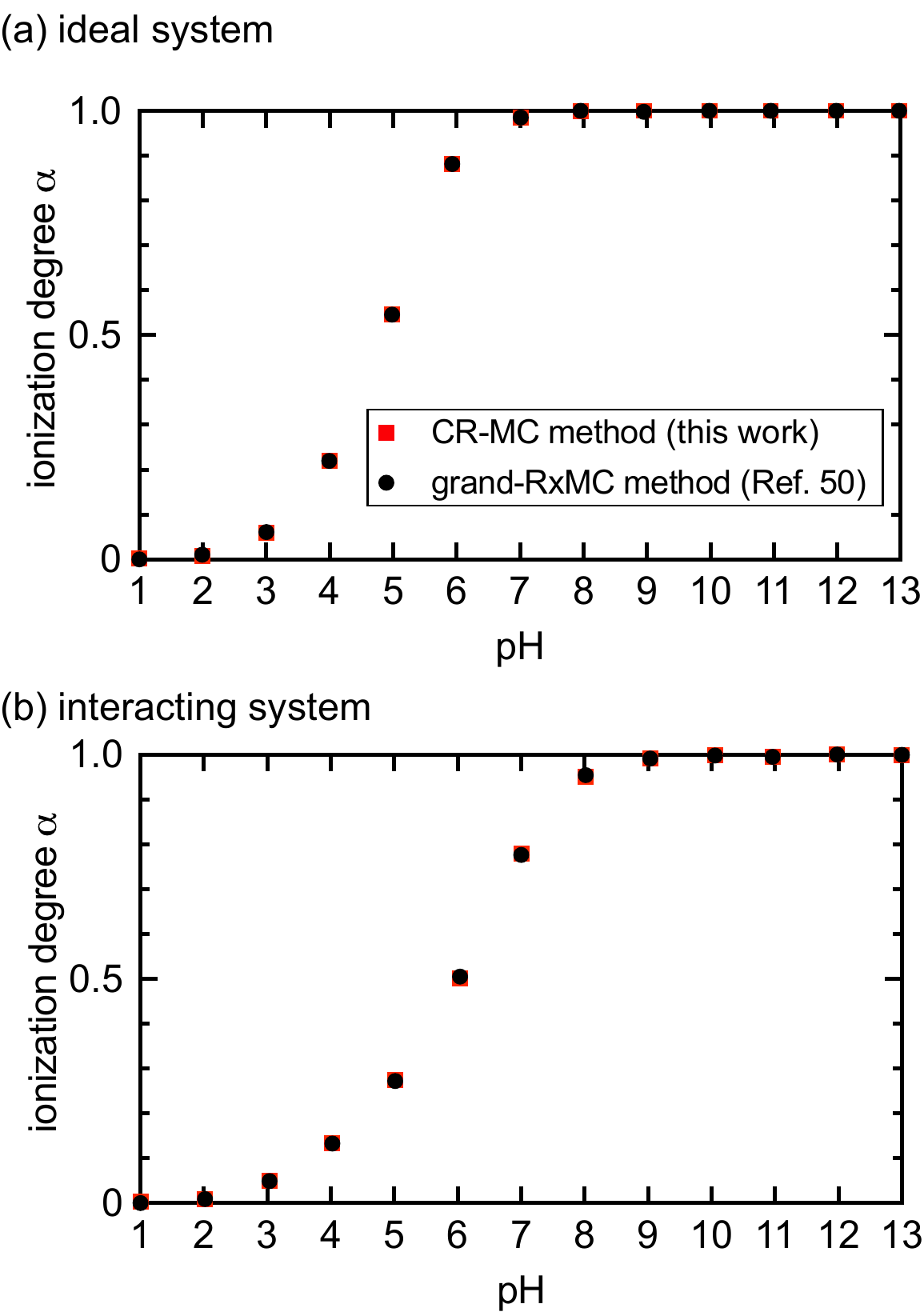}
  \caption{Comparison with previously published results. Average degree of
    dissociation~$\alpha$ obtained using the CR-MC method with the LAMMPS
    implementation (squares) and previously published results that use the
    grand-reaction ensemble method with the ESPResSo package (circles) [Fig.~3
    in Ref.~\onlinecite{landsgesell2020}].  (a)~Ideal system without
    particle–particle interactions.  (b)~Interacting system of polyelectrolyte
    chains.}
\label{fig:Donnan}
\end{figure}

Lastly, we test the CR-MC method and LAMMPS implementation by reproducing
previously published results on acid dissociation (Fig.~\ref{fig:Donnan}).
For this comparison we use dissociation constant $\pKa=4$, salt chemical
potential $\pI_{\tn S^{\pm}}=2$, ion diameter $\sigma=0.355\,\tn{nm}$, Bjerrum
length $l_\tn{B}=2\sigma$, box size $L=29.14l_\tn{B}$, and a total simulation
time of $5\times10^5\tau$ with time step $\delta t=0.005\tau$.  We consider an
ideal system of 800 monomers (Fig.~\ref{fig:Donnan}a) as well as a
polyelectrolyte solution containing 16 polyelectrolyte chains where each chain
contains $N=50$ acid monomers bonded with a FENE potential
(Fig.~\ref{fig:Donnan}b) (see Supporting Information of
Ref.~\onlinecite{landsgesell2020}, Section~S3 for more details). In both cases
we find that our calculation of the average degree of dissociation~$\alpha$ is
statistically identical results to the previously published data.

\newpage

%

\end{document}